\documentstyle[12pt]{article}
\begin{document}

~~~~~~~~~~~~~~~~
\vskip 0.5cm

\begin{center}

{\Large {\bf Experimental limits on the proton life-time from the neutrino
experiments with heavy water}}

\vskip 0.3cm

{\bf V.I.~Tretyak and Yu.G.~Zdesenko}

\vskip 0.3cm

{\it Institute for Nuclear Research, MSP 03680 Kiev, Ukraine}

\end{center}

\begin{abstract}
\noindent
Experimental data on the number of neutrons born in the heavy water targets
of the large neutrino detectors are used to set the limit on the proton
life-time independently on decay mode  through the reaction $d\rightarrow n+?
$. The best up-to-date limit $\tau _p$ $>$ 4$\times $10$^{23}$ yr with 95\%
C.L. is derived from the measurements with D$_2$O target (mass 267 kg)
installed near the Bugey reactor. This value can be improved by six orders
of magnitude with future data accumulated with the SNO detector containing
1000 t of D$_2$O.

~~~~~~~~~~~~~



\end{abstract}

\vskip 0.3cm

\section{Introduction}

The baryon ($B$) and lepton ($L$) numbers are absolutely conserved in the
Standard Model (SM). % due to unbroken global symmetry. 
However, many extensions of the SM, in particular, grand unified theories
incorporate $B$ and $L$ violating interactions, since in the modern gauge
theories conservation of baryon (lepton) charge is considered as approximate
law due to absence of any underlying symmetry principle behind it, unlike
the gauge invariance in electrodynamics which guarantees the massless of
photon and absolute conservation of the electric charge. Therefore, it is
quite natural to suppose the decay of protons and neutrons bounded in
nuclei. The processes with $\Delta B=1$, $\Delta B=2$, $\Delta (B-L)=0$, $%
\Delta (B-L)=2$ have been discussed in literature (see e.g. \cite{Lan81,
Gol80, Kla95} and references therein), while the disappearance of nucleons
(or decay into ''nothing'') has been addressed in connection with possible
existence of extra dimensions \cite{Ynd91, Dva99, Dub00}.

Stimulated by theoretical predictions, nucleon instability has been searched
for in many underground experiments with the help of massive detectors such
as IMB, Fr\'ejus, Kamiokande, SuperKamiokande and others (for experimental
activity see \cite{Kla95,Per84, Bar92} and refs. therein). About 90 decay
modes have been investigated; however, no evidence for the nucleons decay
has been found. A complete summary of the experimental results is given in
the Review of Particle Physics \cite{PDG00}. For the modes in which the
nucleon decays to particles strongly or electromagnetically interacting in
the detector's sensitive volume, the obtained life-time limits are in the
range of $10^{30}-10^{33}$ yr, while for decays to only weakly interacting
products (neutrinos) the bounds are up to 10 orders of magnitude lower \cite
{PDG00, Ber00}. However, because it is not known {\it a priori} which mode
of proton decay (from 90 ones listed in \cite{PDG00}) is preferable, the
limits on the proton decay independent on the channel are very important.

Three approaches were used to establish such limits:

(1) In ref. \cite{Fle58} the bound $\tau (p\rightarrow $?) $>$ 1.3$\times $10%
$^{23}$ yr was determined\footnote{%
We recalculated the value quoted in \cite{Fle58} $\tau(N \rightarrow ?) >3
\times 10^{23}$ yr (given for 232 particles: 142 neutrons and 90 protons)
for 90 protons which should be taken here into consideration ($N$ is $p$ or $%
n$).} on the basis of the limit for the branching ratio of $^{232}$Th
spontaneous fission $\beta _{SF}$. It was assumed that the parent $^{232}$Th
nucleus is destroyed by the strongly or electromagnetically interacting
particles emitted in the proton decay or, in case of proton's disappearance
(or $p$ decay into neutrinos) by the subsequent nuclear deexcitation
process. Using the present-day data \cite{TOI96} on the $^{232}$Th half-life 
$T_{1/2}$ = 1.405$\times $10$^{10}$ yr and limit $\beta _{SF}<$ 1.8$\times $%
10$^{-9}$\%, we can recalculate the value of \cite{Fle58} as $\tau
(p\rightarrow $?) $>$ 1.0$\times $10$^{23}$ yr.

(2) In ref. \cite{Dix70} the limit $\tau (p\rightarrow $?) $>$ 3$\times $10$%
^{23}$ yr was established by searching for neutrons born in liquid
scintillator, enriched in deuterium, as result of proton decay in deuterium (%
$d\rightarrow n+?$).

(3) In ref. \cite{Eva77} the limit $\tau (p\rightarrow 3\nu )>$ 7.4$\times $%
10$^{24}$ yr was determined\footnote{%
The value $\tau (N\rightarrow 3\nu $) $>1.$6$\times $1$0^{25}$ yr quoted in 
\cite{Eva77} as given for 52 particles (28 neutrons and 24 protons) was
recalculated for 24 protons.} on the basis of geochemical measurements with
Te ore by looking for the possible daughter nuclides ($^{130}$Te $%
\rightarrow $ ....$\rightarrow $ $^{129}$Xe), while in refs. \cite{Fir78,
Ste78} the bound $\tau (p\rightarrow 3\nu )>$ 1.1$\times $10$^{26}$ yr was
achieved by the radiochemical measurements with 1710 kg of potassium acetate
KC$_2$H$_3$O$_2$ placed deep underground ($^{39}$K $\rightarrow $ ....$%
\rightarrow $ $^{37}$Ar). In the experiments (2) and (3) both the baryon
number and the electric charge would be not conserved; nevertheless authors
suggested that ''experimenter would be wise not to exclude such processes
from consideration {\it a priori}'' \cite{Eva77}.

The limits \cite{Eva77, Fir78, Ste78} usually are quoted as ''independent on
channel'' \cite{PDG00}, however it is evident that they are valid only for
the proton decay into invisible channels (or disappearance), in which the
parent nucleus is not fully destroyed (like $^{232}$Th in the experiment 
\cite{Fle58}). At the same time, bound on the proton decay from the
deuterium disintegration requires the less stringent hypothesis on the
stability of daughter nuclear system and, hence, it is less model dependent.
Such a limit was established in 1970 \cite{Dix70} and is equal $\tau
(p\rightarrow $?) $>$ 3$\times $10$^{23}$ yr at 68\% C.L.\footnote{%
Because ref. \cite{Dix70} is not the source of easy access, and in ref. \cite
{PDG00}, where this limit is quoted, there is no indication for confidence
level, we suppose that it was established with 68\% C.L.}. This value can be
improved by using the data from the modern neutrino experiments with heavy
water well shielded against cosmic rays and natural radioactivity. With this
aim, in present paper we analyze the measurements of ref. \cite{Ril99} with
the 267 kg D$_2$O target installed at Reactor 5 of the Centrale Nucleaire de
Bugey (France). Further, we show that obtained limit $\tau (p\rightarrow $?)
can be highly improved with the SNO (Sudbury Neutrino Observatory) large
volume detector \cite{Bog00} developed mainly for the Solar neutrino
investigations and containing 1000 t of D$_2$O.

\section{D$_2$O experiment at the Bugey reactor}

The experiment \cite{Ril99} was aimed to measure the cross sections for the
disintegration of deuteron by low-energy electron antineutrinos from nuclear
reactor through reactions $\overline{\nu }_e+d\rightarrow \overline{\nu }%
_e+n+p$ (neutral currents) and $\overline{\nu }_e+d\rightarrow e^{+}+n+n$
(charged currents). Events were recognized by the neutrons they produced.
The detector was located on the depth of 25 meters of water equivalent (mwe)
at 18.5 m distance from the center of the Reactor 5 core at the Bugey site.
The cylindrical target tank, containing 267 kg of 99.85\% pure D$_2$O, was
surrounded by layers of lead (10 cm) and cadmium (1 mm) to absorb thermal
neutrons from external surroundings. The tank with D$_2$O and Pb-Cd shield
was inserted in the center of a large liquid scintillator detector (based on
mineral oil) which served as (inner) cosmic ray veto detector. Subsequent
layer of lead 10 cm thick was aimed to reduce the flux of external $\gamma $
quanta with energies $E_\gamma >2.23$ MeV which can photodisintegrate the
deuterons creating the background events. However this shielding itself was
a significant source of neutrons in the target detector: they were created
due to interaction of cosmic rays with Pb. To suppress this background, an
additional layer of cosmic ray veto detectors was installed outside the Pb
shielding. The outer veto reduced the neutron background in the target by a
factor of near 6. Neutrons were detected by $^3$He proportional counters
installed in the tank with D$_2$O through reaction $^3$He + $n\rightarrow $ $%
^3$H + $p+764$ keV. Further details on the experiment can be found in \cite
{Ril99}.

The decay or disappearance of proton bounded in deuterium nucleus, which
consists only of proton and neutron, will result in the appearance of free
neutron: $d\rightarrow n+?.$ Thus the proton life-time limit can be
estimated on the basis of the neutron rate detected in the D$_2$O volume
when the reactor is switched off. To calculate the $\lim \tau (p\rightarrow $%
?) value, we use the formula 
\begin{equation}
\lim \tau (p\rightarrow ?)=\varepsilon \times N_d\times t/\lim S,
\end{equation}
\noindent where $\varepsilon $ is the efficiency for the neutron's
detection, $N_d$ is number of deuterons, $t$ is the time of measurement, and 
$\lim S$ is the number of proton decays which can be excluded with a given
confidence level on the basis of the neutron background (one-neutron events)
measured in the experiment. Mean efficiency for single neutrons born
isotropically throughout the D$_2$O volume was determined as $\varepsilon
=0.2$9$\pm $0$.01$ \cite{Ril99}. In 267 kg of D$_2$O there is $N_d=1.60$5$%
\times $1$0^{28}$ deuterons. Raw one-neutron rate with the reactor down is
equal 25.28$\pm $0.68 counts per day (cpd) and, corrected for software
efficiency (0.444$\pm $0.005), this rate is $57.00\pm 1.53$ cpd. For very
rough estimate of the $p$ life-time (as the first approximation) we can
attribute all neutron events to proton decays and obtain the $\lim S$~ value
as~59.5 cpd at 95\% C.L. Then, substituting this value in the formula (1) we
get limit $\tau (p\rightarrow $?) $>2.$1$\times $1$0^{23}$ yr with 95\% C.L.%
\footnote{%
The similar limit $\tau (p\rightarrow $?) $>1.$9$\times $1$0^{23}$ yr with
95\% C.L. can be derived from other neutrino deuteron experiment at
Krasnoyarsk (Russia) nuclear reactor \cite{Kra99}.} However, it is evident
that $\tau $ limit derived in this way is very conservative because the
dominant part of observed neutron rate has other origins rather than proton
decay \cite{Bus00,Mik00,Gra00}. On the other hand, if we suppose that all
measured neutron events are belonging to background, then the excluded
number of neutrons due to possible proton decay will be restricted only by
statistical uncertainties in the measured neutron background. Hence, to
estimate value of $\lim S$ we can use so called ''one (two, three) $\sigma $
approach'', in which the excluded number of effect's events is determined
simply as square root of the number of background counts multiplied by one
(two or three) according to the confidence level chosen (68\%, 95\% or
99\%). This method gives us the sensitivity limit of the considered
experiment to the proton decay. Applying it we get $\lim S$ = 3 cpd (at 95\%
C.L.), which leads to the bound\quad $\tau (p\rightarrow $?) $>$ 4$\times $1$%
0^{24}$ yr.

Therefore, we can argue that, with the probability close to 100\%, estimate
of $\tau $ limit is within interval 2$\times $1$0^{23}$ yr --~4$\times $1$%
0^{24}$ yr. In order to fix life-time limit or at least to narrow this
interval, it is necessary to determine the contributions of different
sources to the total neutron rate observed. As it was already mentioned, the
nature and origins of neutron background in neutrino experiments at nuclear
reactors are well known and understood (see, for example, refs. \cite
{Bus00,Mik00,Gra00,Dec99}). The main sources are: (i) interaction of cosmic
muons (escaped an active veto system) with the detector, passive shield and
surrounding materials; (ii) photodisintegration of the deuteron by $\gamma $
quanta (with $E_\gamma >2.23$ MeV), originated from the radioactive
contamination of the detector materials and shield, as well as from
environment pollution; (iii) residual (and non-eliminated by the shield)
neutron background at the nuclear reactor site.

Before coming to details, we would like to remind that crucial
characteristics of any neutrino experiment at reactor are the depth of its
location and distance from the reactor core \cite{Bus00,Mik00,Gra00,Dec99}.
Let us prove this statement by Table 1 with parameters of the most advanced
experiments and by two short citations, from ref. \cite{Bus00}: {\it ''A
striking difference between the two experiments is the amount of overburden,
which may be viewed as the main factor responsible for how the experiments
compare on detector design, event rate, and signal-to-background''}, and
from ref. \cite{Mik00}: {\it ''... a detector should be located sufficiently
deep underground to reduce the flux of cosmic muons - the main source of
background in experiments of this type''}.

\begin{table}[t]
\caption{Main characteristics of the neutrino experiments at nuclear 
reactor}
\begin{center}
\begin{tabular}{|l|l|l|l|l|}
\hline
Experiment & Bugey-1995 & Bugey-1999 & Palo Verde & Chooz \\ 
& \cite{Ach95} & \cite{Ril99} & \cite{Piep99} & \cite{Chooz98} \\ \hline
Depth & 40 mwe & 25 mwe & 32 mwe & 300 mwe \\ \hline
Distance & 15 m & 18.5 m & $\approx $0.9 km & $\approx $1 km \\ \hline
Detector type & Li-loaded & D$_2$O target $+$ & Gd-loaded & Gd-loaded \\ 
& scintillator & $^3$He counters & scintillator & scintillator \\ \hline
Detector mass & $\approx $0.6 t & 267 kg & 11 t & 5 t \\ \hline
Neutron & $\approx $100 cpd/t & $\approx $100 cpd/t & 2.2 cpd/t & 0.24 cpd/t
\\ 
background &  &  &  &  \\ \hline
\end{tabular}
\end{center}
\end{table}

It is clear from the Table 1 that neutron background of different experiments
is decreased as the depth of their location and distance from the reactor
are enlarged. For example, in the Chooz experiment with liquid scintillator
the background was reduced roughly by factor 500 as compared with that of
Bugey ones \cite{Ril99,Ach95} because the Chooz set up was placed deep
underground (300 mwe) and $\approx $1 km away from nuclear reactor \cite
{Chooz98}. We recall that Bugey set ups \cite{Ril99,Ach95} were located only
25--40 mwe overburden (which allows to remove the hadronic component of
cosmic rays but is not enough to reduce the muon flux significantly) and at
15--18 m distance from the reactor core. Thus, the dominant part of neutron
background in \cite{Ril99,Ach95} is associated with the reactor site and
muon flux. Indeed, as it was proved by the detail simulation and careful
analysis of neutron background in reactor-off periods of the experiment \cite
{Ach95}, the 67$\pm $3\% of neutron rate (measured at 15 m distance) are
attributed to known origins (see Table 6 in ref. \cite{Ach95}). On the basis
of comparison of different experiments presented in Table 1, and taking into
account results of background analysis \cite{Ach95}, we can make
semi-quantitative and conservative estimation that at least 50\% of
one-neutron events measured in \cite{Ril99} are caused by the mentioned
sources (i) -- (iii). Hence, attributing remaining part of one-neutron rate
to other unknown background origins, we can accept its value as the excluded
number of proton's decays ($\lim S$ = 30 cpd). Finally, substituting this
number in the formula (1) we obtain

\begin{center}
$\tau (p\rightarrow $?) $>$ 4$\times $10$^{23}$ yr \quad with 95\% C.L.,
\end{center}
\noindent which is higher than previous limit \cite{Dix70}.

\section{Expected improvements with the SNO solar neutrino detector}

The Sudbury Neutrino Observatory (SNO) \cite{Bog00} is an unique large
Cherenkov detector constructed with an emphasis on the study of Solar
neutrinos. The detector, containing 1000 t of 99.917\% isotopically pure
heavy water, is located in the INCO Creighton nickel mine near Sudbury,
Ontario, on the depth of 2039 m (near 6000 mwe); this reduces the muon flux
to 70 muons per day in the detector area. Particular attention is payed to
minimization of radioactive backgrounds. Near 7000 t of ultra-pure light
water shield the central D$_2$O detector from natural radioactivity from the
laboratory walls. All components of the detector are made of selected
materials with low radioactivity contamination.

Solar neutrinos will be detected through the following reactions with
electrons and deuterons: $\nu _i+e^{-}\rightarrow \nu _i+e^{-}$ (elastic
scattering; $i=e,\mu ,\tau $), $\nu _e+d\rightarrow e^{-}+p+p$ (charged
current absorption) and $\nu _i+d\rightarrow \nu _i+n+p$ (neutral current
disintegration of deuteron). Near 9600 photomultiplier tubes are used to
observe the Cherenkov light produced on the D$_2$O volume by high energy
products. Neutrons released in $d$ disintegration will be detected by
neutron capture on deuterons in pure D$_2$O, or by capture on $^{35}$Cl by
dissolving MgCl salt in the heavy water, or by capture on $^3$He using
proportional counters. Further details can be found in \cite{Bog00}.

Extensive Monte Carlo simulations were performed to predict response
functions and numbers of expected events due to interaction of the detector
with Solar neutrinos, natural radioactivity of various detector components,
cosmogenic activities, capture of neutrons, $(\alpha ,p\gamma )$, $(\alpha
,n\gamma )$ reactions outside the SNO detector, etc. (f.e., see \cite{Rob98}%
). Expected number of neutrons from all sources in the D$_2$O volume is
calculated as $\approx $5$\times $1$0^3$ during 1 yr period of exposition,
with main contribution from the Solar neutrinos. Efficiency for $n$
detection is equal 83\% for $n$ capture on $^{35}$Cl \cite{Bog00, Rob98}.

Using these unique features of the SNO detector (super-low background, large
amount of D$_2$O and high sensitivity to neutrons), the limit on the proton
decay independent on channel can be highly improved. Again for very rough
estimate of the $p$ life-time we can conservatively attribute all neutrons
in the D$_2$O volume to proton decays and accept it as the excluded value $%
\lim S$~= 5$\times $1$0^3$ counts. Then, substituting in the formula (1)
values of efficiency $\varepsilon =0.83$, measuring time $t=1$ yr, number of
deuterons $N_d=6\times 10^{31}$ and $\lim S$ = 5$\times $1$0^3$ counts, we
receive

\begin{center}
$\tau (p\rightarrow $?) $>$ 1$\times $1$0^{28}$ yr,
\end{center}

\noindent 
which is about five orders of magnitude higher than present-day limit.
However this value can be improved further by accounting the neutron events
originating from Solar neutrinos and high energy $\gamma $ quanta. Number of
neutrons born in the D$_2$O volume due to disintegration $\nu
_i+d\rightarrow \nu _i+n+p$ can be estimated independently using the
information on the number of Solar neutrino interaction with the detector
volume through neutrino-electron elastic scattering $\nu _i+e^{-}\rightarrow
\nu _i+e^{-}$. Neutrons created by high energy $\gamma $ quanta from natural
radioactivity in the detector components can be also calculated if the
levels of pollution of all materials are measured firmly. In that case the
excluded number of neutrons due to possible proton decay will be restricted
only by statistical uncertainties of the measured neutron background, i.e.
we can apply ''two $\sigma $ approach'' again. It gives $\lim S$ = $2\sqrt{%
5000}$ with 95\% C.L., thus the corresponding bound on the proton life-time
would be equal to

\begin{center}
$\tau (p\rightarrow $?) $>$ 4$\times $1$0^{29}$ yr\quad with 95\% C.L.,
\end{center}

\noindent which can be considered as the maximal sensitivity of the SNO
detector for the proton decay independent on channel.

\section{Conclusion}

The data of the Bugey experiment \cite{Ril99}, aimed to measure the cross
sections for the deuteron disintegration by antineutrinos from nuclear
reactor, were analyzed to set the proton life-time limit. The obtained value 
$\tau (p\rightarrow $?) $>$ 4$\times $1$0^{23}$ yr at 95\% C.L. is higher
than the limit established in the previous experiment \cite{Dix70}. With the
future data from the SNO Solar neutrino detector (containing 1000 t of heavy
water) life-time limit will be improved up to the value $\tau (p\rightarrow $%
?) $>$ 4$\times $10$^{29}$ yr, which is, in fact, close to the bounds
established for the particular modes of the nucleon decays to charged or
strongly interacting particles and would be of a great importance for many
extensions of the modern gauge theories.

\end{document}